# Growth of delafossite $CuAlO_2$ single crystals in a reactive crucible

*Du hyung Kim[1], Minsik Kong[1], Myeongjun Kang[1], Minjae Kim[1], Seohee Kim[1], Youngwook Kim[2], Sangmoon Yoon[3], Jong Mok Ok,[1,]\**

[1]Department of Physics, Pusan National University, Busan 46241, Korea

[2]Department of Physics and Chemistry, DGIST, 42988 Daegu, Korea

[3]Department of Physics, Gachon Universtiy, 13120 Seongnam, Korea

*Correspondence should be addressed to okjongmok@pusan.ac.kr

**ABSTRACT.**

Delafossite oxide $CuAlO_2$ has engaged great attention as a promising *p*-type conducting oxide. In this work, high-quality $CuAlO_2$ single crystals with a size of several millimeters (mm) are successfully achieved with a *reactive* crucible melting method. The crystals are characterized by X-ray diffraction, scanning electron microscopy with energy-dispersive spectroscopy, transport measurement, and magnetic susceptibility measurement. The grown single crystals are free of contamination from a copper oxide flux. This work provides a new approach to growing high-quality delafossite oxide $CuAlO_2$ with a few mm size.

**INTRODUCTION.**

Delafossite CuAlO$_2$ (CAO) is highly attractive as a *p*-type metal oxide owing to its excellent conductivity compared to other *p*-type semiconductors [1, 2]. Moreover, CAO exhibits high transmittance of up to 70% in a thin regime [3,4,5], drawing more attention to its potential application for optoelectronic devices. Mainly, CAO has been synthesized in thin films and polycrystalline powders, where grain boundaries predominantly determine transport and optical properties. Meanwhile, a slow-cooled method has been utilized to obtain CAO single crystals (described in Ref 7). In this method, a complex procedure is required to eliminate the copper oxide Cu$_2$O flux; for example, the single crystals are put in HNO$_3$ at 100°C for four days to resolve the remaining flux [7]. Therefore, the development of a facile method to grow high-quality and large-sized CAO single crystals is demanded to investigate their intrinsic properties further and advance their applicability to optoelectronic devices.

According to the literature that reports on the Cu-Al-O phase diagram, a CuAlO$_2$ phase would be obtained from mixed starting reagents of Cu$_2$O with Al$_2$O$_3$ at high temperature (>1100 °C)[8]. Furthermore, Cu$_2$O powders are easily evaporated at high temperatures, which may promote the reaction between the vaporized Cu$_2$O flux and the Al$_2$O$_3$ crucible. In this context, the vaporized (or melted) Cu$_2$O flux can react to the surface of the alumina crucible, producing the CAO phase through the following reactions: Cu$_2$O (flux) + Al$_2$O$_3$ (crucible) → 2 CuAlO$_2$ (crystal). In this study, we report a facile method to synthesize high-quality CAO single crystals on the millimeter (mm) scale. Cu$_2$O powders are solely heated in the alumina crumble at high temperatures; then, Cu$_2$O powders react with Al$_2$O$_3$ crucible, resulting in mm-sized CAO single crystals. We confirmed that the CAO single crystals are paramagnetic semiconductors, which is in good agreement with other crystals grown by the conventional slow-cooling method.

**EXPERIMENTS**

High-quality mm-sized CAO single crystals were grown by the deliberate reaction between $Cu_2O$ flux and $Al_2O_3$ crucible. Only $Cu_2O$ powder (0.5 – 2.5 g) was put into the alumina crucible, which was not covered by a lid. The lidless crucible allows the supersaturation state of CAO in the evaporating $Cu_2O$ flux. The crucible was located inside a box furnace and heated to 1225°C. After holding the temperature at 1225°C for 12h, the furnace was cooled at a rate of 0.5 − 1°C/h to 1150°C, then cooled to room temperature by switching off the power. Many shiny and thin hexagonal plate-like CAO crystals were obtained. The obtained crystals were mechanically isolated from the alumina crucible. CAO single crystals are usually 1mm x 1mm. The crystal structure of CAO was examined by X-ray diffraction (XRD) with Cu Kα radiation (Bruker D8 Discover X-ray Diffractometer installed in Quantum Matter Core-Facility of Pusan National University). The structural analysis was performed with the help of Full-Prof Suite software. The chemical composition was obtained by scanning electron microscopy with energy-dispersive spectroscopy (SEM/EDS). Temperature dependence of the resistivity and magnetic susceptibility were measured with a vacuum probe station and a magnetic properties measurement system at Pusan National University.

**RESUTS & DISCUSIONS**.

Figures 1(a)-(d) illustrate the single crystal growth process. At high temperatures, CAO phases were synthesized at the interface of $Cu_2O$ flux and alumina crucible by the chemical reaction. At the same time, some of the flux is evaporated, generating a supersaturated solution of CAO and $Cu_2O$. Figures 1(a)-(b) describe the process of supersaturation, in which the concentration of CAO (black) increases. Since supersaturation is the driving force of the crystallization process, the CAO crystals form naturally, as shown in Figures 1(c)-(d). Figure 2(e) indicates the cross-section of the crucible after the thermal process. The enlarged figure is shown in Figure 1(f). We found that $Cu_2O$ flux penetrated the crucible and turned black, creating dents on the crucible surface. The reacted surface of the crucible provides direct evidence that the $Cu_2O$ flux reacts with the alumina crucible from the CAO phase.

In order to obtain large-sized CAO single crystals, we optimized the growth conditions as summarized in Table 1. First, crystal growth using various amounts of $Cu_2O$ flux has been attempted (growths A & B). Substantial size single crystals of CAO were obtained when a sizable amount of $Cu_2O$ flux was employed. However, it was difficult to separate the crystals in this case due to a massive amount of remaining flux. Another critical parameter we discovered is the size of the alumina crucible. We utilized two crucibles with different areas (growths B & C). The CAO crystals grown in the crucible with a small area are larger than those grown in a wide crucible. The $Cu_2O$ flux evaporates rapidly in a wide crucible, so the crystal cannot be large enough. On the other hand, in the narrow crucible, the evaporation of the $Cu_2O$ flux is significantly suppressed, allowing the flux and the crucible to react sufficiently to create large-sized crystals. $Cu_2O$ residues were found particularly in the narrow crucible as a side effect (growth C). These results suggest that optimizing the amount of flux and the size of a crucible is essential for obtaining large-sized CAO single crystals.

The mm-sized CAO single crystals were successfully obtained in the optimized growth condition (growth D). The thermal process was repeated twice to ensure sufficient time for the CAO crystals to make mm-size. The macroscopic size of the grown CAO single crystals is as large as ~1 mm × 1 mm with shiny black color, as shown in the optical image (see Figure 2(c)).

The crystals have triangular or hexagonal plate shapes, unlike cubic like $Cu_2O$ single crystals [9], supporting that CAO single crystals were successfully obtained from the $Cu_2O$ flux. The XRD result supports the single crystalline quality of the CAO single crystal, as shown in Figure 2(a). The XRD shows only (00$L$) reflections, implying that the growth direction of the CAO crystal is along the $c$-axis. The crystal structure of the CAO single crystal was determined by the powder XRD and Rietveld refinement, as shown in Figure 2(b). The CAO powder was prepared by grinding single crystals. The experimentally obtained XRD patterns (red dots) agree well with the calculated XRD pattern (solid black line). The lattice parameters $a$, $b$, and $c$ estimated from the Rietveld refinement are $a=b=2.861(7)$Å, $c=16.974(4)$Å, in good agreement with references [10]. More specific parameters obtained from the refinement are shown in Table 2. The elemental stochiometric composition of the obtained single crystal was determined by SEM/EDS technique. Only Cu, Al and O are detected, confirming absence of impurities. The chemical composition is uniform with a ratio of Cu:Al = 1.06:1. Taking into account the accuracy of SEM/EDX (typical error $\pm$ 5%), this results show that nearly stoichiometric crystals can be achieved by growth in the reactive crucible.

Figure 3(a) displays the temperature dependence of resistivity $\rho(T)$ measured in the temperature range from 300 to 450 K. It is worth noting that reliable and reproducible measurements with low contact resistance were particularly limited to the high-temperature range ($T > 300$ K). Our crystal exhibits semiconductor behavior. The resistivity drops from ~123 $\Omega$cm at 300 K to ~ 1 $\Omega$cm at 450 K. Figure 3(b) shows that the plot of log($\rho$) versus 1000/$T$ is linearly dependent. Such a linear dependence can fit an Arrhenius- type formula $\rho = \rho_0\, e^{U/kT}$, where $\rho_0$ is the prefactor of resistivity, $U$ is the activation energy, and $k$ is the Boltzmann constant [11]. The activation energy is estimated to be $U = 340$ meV, which is consistent with that expected from the temperature dependence of the carrier density [12]. The activation energy is smaller than the bandgap $E_g \sim 3.5$ eV[4], confirming that CAO is an extrinsic *p*-type semiconductor with shallow electron acceptors; Cu vacancies are reported to act as shallow electron acceptors in CAO[3].

Figure 4 presents the temperature dependence of magnetic susceptibility $\chi(T)$ measured under magnetic field $H = 1$kOe. Our CAO single crystals show a paramagnetic behavior, as reported in early studies [12-14]. Stoichiometric CAO without point defects is supposed to exhibit diamagnetism because the electrons in $Cu^+$, $Al^{3+}$, and $O^{2-}$ are all paired [12]. In reality, CAO exhibits weak paramagnetism, which could be attributed to $Cu^{2+}$ with $d^9$ orbital occupations. As the valence-band top of CAO consist of Cu *d* orbitals, hole carriers directly influence the oxidation state of $Cu^+$. It is worth noting that the magnetic susceptibility of our crystals ($\chi(5\text{ K}) = 8.5 \times 10^{-6}$ emu/g·Oe) is lower than that of polycrystalline powders ($\chi(5\text{ K}) = 2 \times 10^{-5}$ emu/g·Oe)[15], suggesting that the density of Cu vacancy in this sample is lowered than that in previously synthesized samples (see Figure 4(b) for comparison).

For further understanding, $\chi(T)$ was fitted to the modified Curie-Weiss law: $\chi = \chi_0 + \frac{C}{T-\theta}$ [16]. Here $\chi_0$ is the temperature independent contribution from diamagnetism and Van Vleck susceptibility, $\theta$ is the Curie-Weiss temperature, $C = Ng^2\mu_B^2 S(S+1)/3k_B$ is the Curie constant with number of magnetic ions per gram $N$, Boltzmann constant $k_B$, Lande g-factor g, magnitude of electron spin S, and Bohr magneton $\mu_B$. The best-fit values to $\chi_0, \theta, C$ are $1.07 \times 10^{-6}$ emu Oe$^{-1}$cm$^{-3}$, -0.31K, and $2.23 \times 10^{-4}$ emu K Oe$^{-1}$cm$^{-3}$, respectively. These values are coherent with those obtained for single crystals grown by the slow-cooling method [12]. Assuming that the paramagnetic moment emanates from the $Cu^{2+}$ defect states with S=1/2, the density of paramagnetic defect is estimated to be $N = 3.58 \times 10^{20}$ cm$^{-3}$, which is about 1.4 % of that of Cu sites in CAO ($2.5 \times 10^{22}$ cm$^{-3}$). We note that the estimated ratio of defect states is smaller than the polycrystalline powders (~2.8 – 3.1 %) [14] and comparable to single crystals synthesized by a slow-cooled method (~1.3 %) [12]. The $\chi_0$ also arises presumably due to Cu$^{2+}$ ions. While the core electron shells of Cu$^{1+}$, Al$^{3+}$, and O$^{2-}$ contribute to diamagnetic susceptibility, as discussed in a previous report [12], Van Vleck contribution can arise from the open shells of Cu$^{2+}$ ions [17].

**CONCLUSION.**

In summary, we have reported the growth of mm-sized CAO single crystals in a reactive alumina crucible. Unlike the conventional flux growth, in which the flux acts as a solvent to dissolve the desired substance, in the method reported in this paper, the flux also plays the role of the starting material that reacts with reactive crucible [18]. High-quality single crystals of CAO without the remaining flux were successfully obtained by the reaction between the Cu$_2$O flux and alumina crucible. The structural, transport, and magnetic properties of our CAO single crystal were investigated. CAO single crystals show semiconducting and weak paramagnetic behavior, which is in good agreement with other crystals grown by the conventional slow-cooling method. This growth method we reported here can eliminate the process of removing residuals, which can solve many problems caused by the residuals, thus facilitating relevant studies on the high-performance $p$-type conducting oxides.

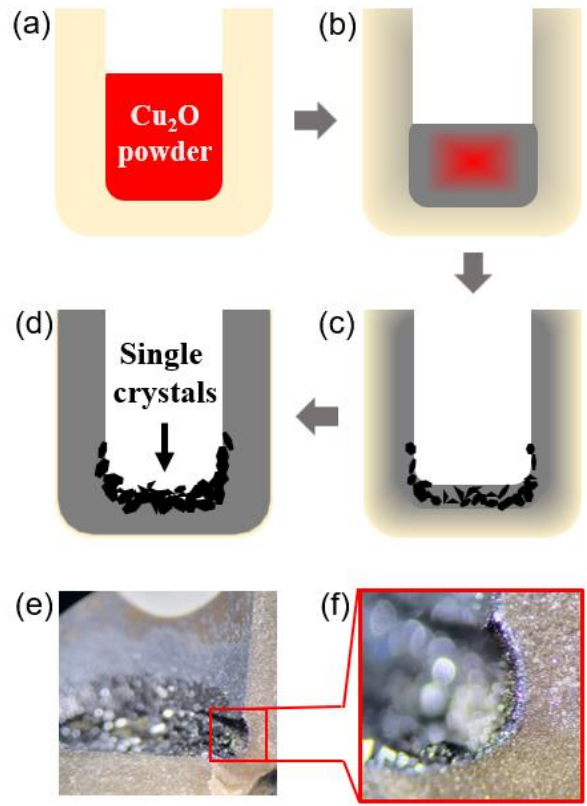

**Figure 1 Schematics of single crystal growth of CuAlO$_2$ in a reactive alumina crucible.** (a) The alumina crucible contains only Cu$_2$O flux powder. (b) The Cu$_2$O flux reacts with the surface of the alumina crucible. (c) During the cooling process, small-sized single crystals form on the surface of the alumina crucible. (d) Large-sized single crystals are obtained after the thermal process. (e) A cross-sectional image of the alumina crucible after thermal process. (f) An enlarged cross-sectional image of the reacted surface of the alumina crucible.

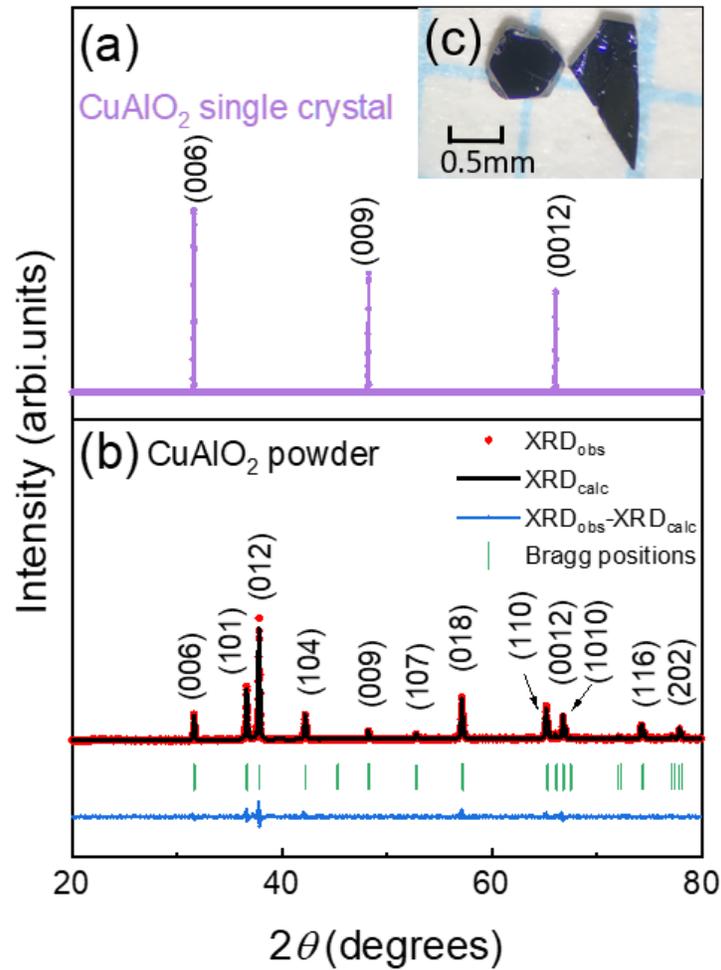

**Figure 2. X-ray diffraction of CuAlO$_2$ single crystal grown in a reactive alumina crucible.** (a) X-ray diffraction pattern CuAlO$_2$ single crystal oriented along the *c*-axis. The observed patterns suggest that the *c*-axis is perpendicular to the surface. (b) Rietveld refinement of the X-ray diffraction pattern of the grounded single-crystals of CuAlO$_2$. Red circles, solid black line, and blue line show the observed intensity, the calculated pattern, and the difference between the observed and calculated patterns. Vertical green bars show the Bragg peak's positions. (c) An optical image of CuAlO$_2$ single crystals.

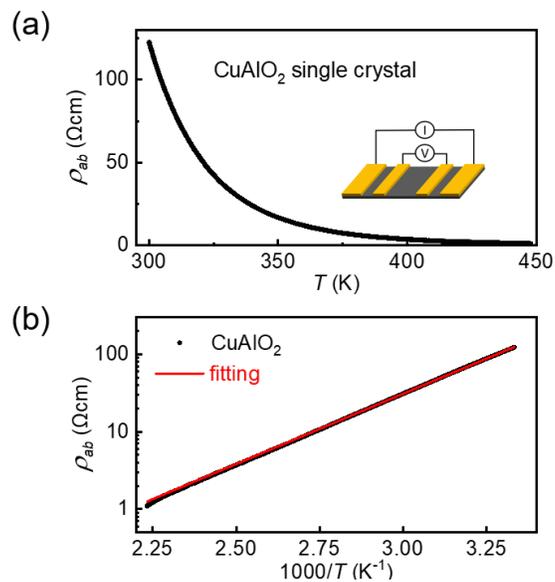

**Figure 3. Transport properties of CuAlO₂ single crystal.** (a) Temperature dependencies of resistivity of CuAlO$_2$ single crystal from 300 to 450K measured in the in-plane direction. Insert shows the contact configuration (4-probe method). (b) Arrhenius plot for the determination activation energy of CuAlO$_2$. Solid black and red lines show the measured data and fitted curve, respectively.

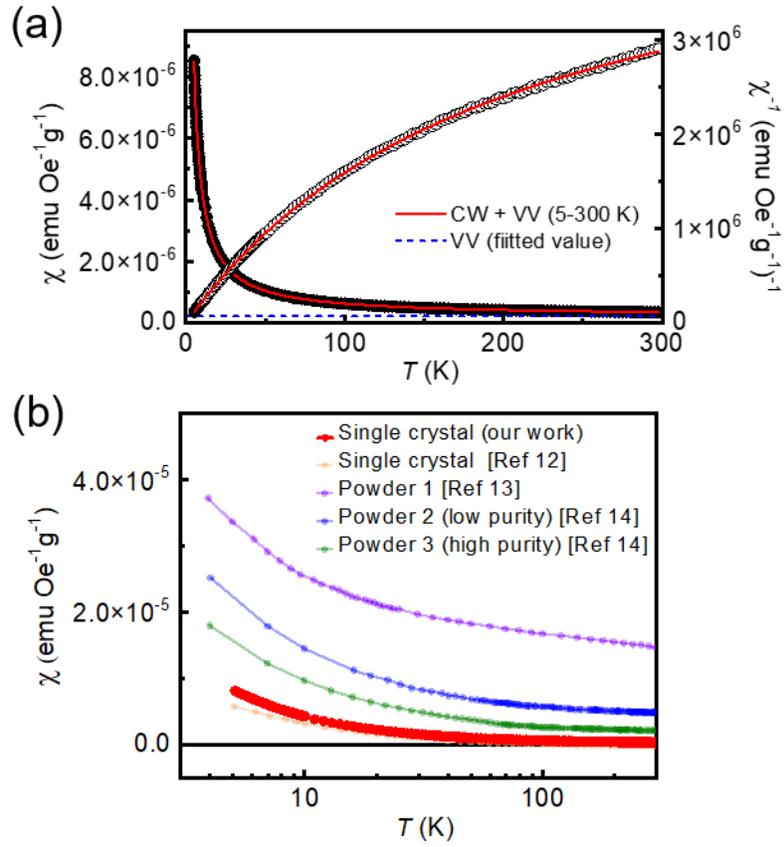

**Figure 4. Magnetic susceptibility of CuAlO$_2$ single crystal.** (a) Temperature dependence of magnetic susceptibility $\chi(T)$ (left axis, filled symbol) and inverse magnetic susceptibility $\chi^{-1}(T)$ (right axis, open symbol) of CuAlO$_2$ single crystal measured at an applied magnetic field of 1 kOe in the temperature range from 5 to 300K. Measured (*T*) (black dots) is well-fitted by the Curie-Weiss law plus a Van Vleck (VV) contribution (red solid line). (b) $\chi(T)$ of our CuAlO$_2$ single crystal (red dots) compared to that of previous reports [12-14].

| Sample | Cu$_2$O [g] | Bottom area [mm$^2$] | Number of thermal process | Size of the crystals [mm] |
|---|---|---|---|---|
| A | 2.5 | 1485 | 1 | 0.4 |
| B | 0.5 | 1485 | 1 | 0.2 |
| C | 0.5 | 50 | 1 | 0.5 |
| D | 2.5 | 227 | 2 | 1.9 |

**Table 1.** Summary of growth conditions.

| Atom position | x | y | z | B$_{iso}$ | Occ |
|---|---|---|---|---|---|
| O | 0 | 0 | 0.10866 | 6.72100 | 0.27871 |
| Cu | 0 | 0 | 0 | 4.50954 | 0.09486 |
| Al | 0 | 0 | 0.49129 | 4.40054 | 0.10902 |

**Table 2.** Detailed structural information on CuAlO$_2$ single crystal obtained from Rietveld refinement parameters. The space group and lattice parameters are R$\bar{3}$m, $a=b=2.861(7)$Å, $c=16.974(4)$Å, and $V=120.3868$ Å$^3$, respectively.